\documentclass[twocolumn,prd,aps,showpacs,amsmath,amssymb]{revtex4-1}
\usepackage{graphics}
\usepackage{graphicx}
\usepackage{dcolumn}
\usepackage{bm}
\usepackage{mathrsfs}
\usepackage{pstricks}
\usepackage{color}
\usepackage{slashed}
\usepackage{amsmath}
\usepackage{epsfig}
\usepackage{amsfonts}
\usepackage{amssymb}
\usepackage{amsthm}

\def\beq{\begin{equation}}
\def\eeq{\end{equation}}
\def\bea{\begin{eqnarray}}
\def\eea{\end{eqnarray}}

\def\Re{\textrm{Re}}

\begin{document}

\title{Riemann zeta zeros \\ and \\ zero-point energy}

\author{J. G. Due\~nas}
\email{jgduenas@cbpf.br}
\affiliation{Centro Brasileiro de Pesquisas F\'{\i}sicas, Rio de Janeiro, RJ 22290-180, Brazil}

%

\author{N. F. Svaiter}
\email{nfuxsvai@cbpf.br}
\affiliation{Centro Brasileiro de Pesquisas F\'{\i}sicas, Rio de Janeiro, RJ 22290-180,
Brazil}

\begin{abstract}

We postulate the existence of a self-adjoint operator associated to a system with countably infinite
number of degrees of freedom whose spectrum is the sequence of the nontrivial zeros of the Riemann zeta function.
We assume that it describes a massive
scalar field coupled to a background field in a $(d+1)$-dimensional flat space-time. The scalar field is confined to the interval $[0,a]$ in one dimension and
is not restricted in the other dimensions.
The renormalized zero-point energy of this system is presented using techniques of dimensional and analytic regularization.
In even dimensional space-time, the series that defines the regularized vacuum energy is finite. For the odd-dimensional case,
to obtain a finite vacuum energy per unit area we are forced to introduce mass counterterms.
A Riemann mass appears, which is the correction to the
mass of the field generated by the nontrivial zeros of the Riemann zeta function.

\end{abstract}



\maketitle

\section{Introduction}

\quad $\,\,$ The Riemann hypothesis states that the complex zeros of the Riemann zeta function $\zeta(s)$ must all lie on
the critical line $\Re(s)=1/2$
\cite{riem,titchmarsh,borwein,rev}. This conjecture is an unsolved problem in mathematics and
all efforts to prove it have remained unsuccessful.
Hilbert and P\'olya suggested that there might be a spectral interpretation of the
nontrivial zeros of the Riemann zeta function, which could be the
eigenvalues of a self-adjoint operator in an appropriate Hilbert space.

The work of Bohigas and Giannoni \cite{bohigas} and also Berry \cite{berry}, showed that the quantum energy spectra of a
classically chaotic systems exhibit universal spectral correlations which are described by random matrix theory \cite{mehta}.
Numerical evidences indicate that the distribution of spacing between nontrivial zeros of the Riemann zeta
function is statistically identical to the distribution of eigenvalue spacings in a Gaussian unitary ensemble
of random matrices. Such a conjecture prompted many authors to consider the
Riemann hypothesis in the light of random matrix theory and quantum mechanics of classically
chaotic systems ~\cite{berry3,berry4,bu2,bourgade,sierra2,sierra1}.
These facts have also led some authors to consider a
Fermi gas with a fully chaotic classical dynamics~\cite{le,le2}. These authors considered the imaginary part
of the complex zeros of the Riemann zeta function as the single-particle levels of a fermionic many-body system,
known in the literature as the Riemannium.

A crucial question is whether it makes sense to suppose that this elusive operator, conjectured by
Hilbert and P\'olya occurs in a quantum  mechanical system with countably infinite number of degrees of freedom.
Here we are interested in
investigating the consequences of the existence of such
operator in a quantum field theory framework. The assumptions are:
the operator must be defined in a bounded region of space, since the spectrum is discrete,
acts on scalar functions defined in flat space-time, and finally,
is self-adjoint in some Hilbert space.

In the approach of quantum field theory using functional methods \cite{fu1,fu2},
Gaussian path integrals yield expressions, which depend on the differential operator's determinant. Although these determinants diverge,
finite regularized values can be obtained using the spectral zeta function regularization~\cite{se,ray,dowker,ha,vo,vor,quine,elizalde,dune}.
Using that the prime zeta function can be analytically extended only in the strip
$0<\sigma\leq 1$ \cite{lan,carl}, Menezes and Svaiter \cite{ga} concluded that the sequence of prime numbers cannot be associated with some
hypothetical linear operator of a physical system with infinitely many degrees of freedom. This result was generalized by Andrade to other sequences of numbers
motivated by number theory ~\cite{andrade}.
Later, Menezes, Svaiter, and Svaiter  first considered the same situation with numerical sequences whose asymptotic distributions are not
``far away" from the asymptotic distribution of prime numbers \cite{gn2}. Next, using the construction of the so-called
super-zetas or secondary zeta functions built over the Riemann zeros, i.e.,
the nontrivial zeros of the Riemann zeta function~\cite{gui,delsarte,cha1,cha2,ivic,voros1,superzeta}, and the regularity properties of one of these secondary zeta
functions at the origin, these authors have
shown that the sequence of the nontrivial zeros of the Riemann zeta function can in
principle be interpreted as being the spectrum of a self-adjoint operator acting on scalar fields of some hypothetical system.

In the functional approach to Euclidean scalar field theory there are three fundamental objects to be considered. The generating functional of all
Schwinger functions $Z[J]$, the generating functional of connected Schwinger functions $W[J]$,  and finally the generating functional of proper vertices $\Gamma[\varphi]$.
Since in perturbation theory the proper vertices are given by the sum of one-particle irreducible diagrams, $\Gamma[\varphi]$ is the generating functional of
the sum of one-particle irreducible Green's functions. Performing a loop expansion we can define the one-loop effective action \cite{ea}.
Since, there is a relationship between the Casimir energy and the one-loop effective
action \cite{casimir,milton,plunien,mostepanenko,Bordag,blau},
it is natural to enquire whether is it possible to calculate the renormalized vacuum energy associated with a hypothetical system where the imaginary part of
the complex zeros of the zeta function appear in the spectrum of the vacuum modes.

The aim of this paper is to discuss the renormalization of the vacuum expectation value of the energy operator, i.e., the renormalized zero-point energy
associated to a massive scalar field defined in a $(d+1)$-dimensional flat space-time,
assuming that one of the coordinates lies in an finite interval $[0,a]$.
We call such configuration a slab-bag \cite{nn}. We consider a
linear operator $-\Delta_{d-1}-{M}$, where $\Delta_{d-1}$ is the usual Laplacian defined in a $(d-1)$-dimensional space and ${M}$
is an unknown operator such that its eigenvalues are the imaginary part of the complex zeros of the Riemann zeta function.
We are using the same idea of the papers \cite{actor,caruso,matt}, where some confining potential acts as a pair of effective plates.
One way to implement our model is to assume that the scalar field is coupled to a confined background
field $\sigma(\textbf{x}_\perp,z)$ with an
interaction Lagrangian ${\cal{L}}_{int}=\sigma(\textbf{x}_\perp,z)\,\varphi^{2}(t,\textbf{x}_\perp,z)$ such
that the system is confined in the interval $[0,a]$ in one dimension and
is unrestricted in the other spatial directions. The idea of substituting the hard boundary conditions on some surface by
potentials was used by many authors.
For instance, the zero-point energy of a quantum field as the
limit of quantum field theory coupled to a background was used in Ref.
\cite{jaffe1,jaffe2,jaffe3}.

The organization of this paper is the following.
In section II we discuss arithmetic gases, where the sequence of prime numbers is used in a quantum field theory framework.
In Section III, we study a massive scalar field in a $(d+1)$-dimensional space-time,  for which the
coordinate $x_d$ lies in a finite interval $[0,a]$,
assuming that in the spectrum of the vacuum modes appear the nontrivial zeros of the
zeta-function. In section IV the renormalized zero-point energy of this system is presented using a combination between
dimensional and analytic regularization procedures.
Conclusions are given in Section V. In Appendix A we present the analytic extension for one of the super-zeta functions
following the classical paper of Delsarte. In Appendix B we show that the renormalized vacuum energy is identified with the
constant term in the asymptotic expansion of the regularized vacuum energy. In the paper we use $k_{B}=c=\hbar=1$.

\section{Arithmetic Riemann gas in quantum field theory}

In this section we discuss how is the sequence of prime numbers has been used in quantum field theory framework.
As a tool to show connections between number theory and physics using statistical-mechanics methods,
some authors introduced the sequence of prime numbers in quantum field theory~\cite{stn,sss,bakas,julia,sp,spector}.
Let us consider a non-interacting bosonic field theory with
Hamiltonian
\begin{equation}
H_B=\omega\sum_{k=1}^{\infty}\ln(p_{k})b^{\dagger}_{k}b_{k},
\label{25}
\end{equation}
where $b_k^{\dagger}$ and $b_{k}$ are respectively the creation and annihilation operators of quanta associated
to the bosonic field and $p_{k}$ are the sequence of prime numbers. We have $p_{1}=2,\,p_{2}=3,...$ and so on.
The energy of each mode is given by $\nu_{k}=\omega\ln\,p_{k}$.
The arithmetic gas partition function is exactly the Riemann zeta function, i.e.,
$Z=\zeta(\beta\omega)$. Since the Riemann zeta function has a simple pole at $s=1$, there is a Hagedorn temperature
above which the system can not be heated up \cite{hagedorn}. Other property of such system is that the
asymptotic density of states increases exponentially with energy. The Riemann gas is also called an arithmetic gas.

For a free arithmetic fermionic gas the Hamiltonian is given by
\begin{equation}
H_F=\omega\sum_{k=1}^{\infty}\ln(p_{k})c^{\dagger}_{k}c_{k},
\label{26}
\end{equation}
where $c_k^{\dagger}$ and $c_{k}$ are respectively the creation and annihilation operators of quanta associated
to the fermionic field and the $p_{k}$ are again the sequence of prime numbers. Making use of the arithmetical M\"obius function \cite{hardy}
defined by $\zeta^{-1}(s)=\sum_{n=1}^{\infty}\mu(n)/n^{s}$
it is possible to show that the partition function of the fermionic system is given by $\zeta(\beta\omega)/\zeta(2\beta\omega)$.
The M\"obius coefficients are $\mu(1)=1$ and $\mu(n)=0,1$ or $-1$, depending whether, $n$ is divisible by a square of a prime number,
or is a product of an even number of primes, all
different, or of an odd number of primes, all different respectively.
As discussed in the literature, from the thermal partition function associated to the Hamiltonians $H_{B}$ and $H_{F}$ we get
\begin{equation}
Z_{F}(\beta)\,Z_{B}(2\beta)=Z_{B}(\beta).
\end{equation}
The noninteracting mixture of two systems, each of its own temperature $\beta^{-1}$ and $(2\beta)^{-1}$ is equivalent to another bosonic system with
temperature $\beta^{-1}$.

Let us show that a alternative bosonic Hamiltonian similar to $H_B$ can not be constructed with the sequence of prime numbers or a sequence of the power of them in the spectrum. Only the $\omega\ln p_{n}$ spectra is allowed.
To proceed let us use the the following representation for $\ln x=\lim_{s\rightarrow 0}\frac{x^{s}-1}{s}$, the replica method used to study disordered systems \cite{glass,do}.
The Hamiltonian of this non-interacting bosonic field theory can be written as
\begin{equation}
H=\lim_{s\rightarrow 0}\omega\biggl(\sum_{k=1}^{\infty}\frac{p_{k}^{s}}{s}b^{\dagger}_{k}b_{k}-\sum_{k=1}^{\infty}\frac{b^{\dagger}_{k}b_{k}}{s}\biggr).
\label{cas}
\end{equation}
Although the Hamiltonian of the Riemann gas is obtained only when ${s\rightarrow 0}$, we can ask if there
is a Hamiltonian for generic value of $s$. The case with $s=1$,  is exactly a system with the spectrum proportional to the prime numbers.
In the following we use the results obtained in \cite{ga} to prove that only the Riemann gas is well defined. Using the
definition of the prime zeta function $P(s)$, $s=\sigma+\tau i$, for
$\sigma,\tau \in  \mathbb{R}$, given by

\begin{equation}
P(s)=\sum_{\left\{p\right\}}\,p^{-s},  \,\,\Re(s)>1,
\label{21}
\end{equation}
where the summation is performed over all primes \cite{carl,lan}, it was shown that the free energy of a system with the
sequence of prime numbers as the spectrum does not exist.
This is related to the fact that there is no analytic extension for $P(s)$ to the half-plane $\sigma\leq 0$.

The same argument can be used to prove that there is no fermionic system with infinitely number of degrees of freedom
whose spectrum is composed by the sequence of prime numbers. Let us consider Dirac fermions interacting with an external
field $A_{\mu}(x)$ in Euclidean space.
The fermionic path integral is not well defined since the determinant
of the Dirac operator diverges because it is an unbound product of increasing eigenvalues $\lambda_{n}$.
It is possible to define a regularized determinant using the spectral zeta function associated with the Dirac operator and
the principle of analytic continuation.
The analytic continuation of the spectral zeta function must be regular at the origin, i.e., $s=0$. The
eigenvalues of the Dirac operator cannot be the sequence of prime numbers by the same reason we discussed for the bosonic case.

Although there is a reciprocity between the set of prime numbers and nontrivial zeros of the zeta function,
these two sequences of numbers have totally distinct behavior with respect to being the spectrum of a
linear operator associated to a system with countable infinite number of degrees of freedom.
In the next section we assume that there is a spectral interpretation of the
nontrivial zeros of the Riemann zeta function. The nontrivial zeros are the
eigenvalues of a self-adjoint operator in an appropriate Hilbert space. We are interested in studying the consequences of the existence of such
operator in a quantum field theory framework.

\section{The zeta zeros and the vacuum energy}

Here we are interested in studying finite size systems \cite{b1,b2,b22,b3}, where the
translational invariance is broken and the imaginary part of the complex zeros of the Riemann zeta function appear as the
spectrum of ${M}$. We refer as finite size system to any system that has finite size in at least on space dimension.
We use cartesian coordinates $x^{\mu}=(t,\textbf{x}_\perp,z)$.

Let us consider the complex zeros of the zeta function
 $\rho=\beta+i\gamma$ for
$\beta,\gamma \in  \mathbb{R}$ and $0<\beta<1$.
The Riemann hypothesis is the conjecture that $\beta=1/2$.
We will assume through the paper the validity of this conjecture and also take only the positive non-trivial zeta zeros, i.e., $\gamma>0$.
For the prime numbers, using the
prime number theorem, we get the asymptotic regimes $p_{n}\sim\,n\ln n$.
Let the zeros of $\zeta(s)$ on the line $Re(s)=1/2$ be $\rho_k=1/2+i\gamma_k$ $(k=\pm1,\pm2, ...)$ with $\gamma_{-k}
=-\gamma_{k}$ and $\gamma_{0}=0$. Let us assume that these zeros are simple.
If the zeros with $\gamma>0$ are arranged in a sequence $\rho_{k}=1/2+i\gamma_{k}$ so that
$\gamma_{k+1}>\gamma_{k}$, then $|\rho_{k}|\sim\gamma_{k}\sim 2\pi k/\ln k$ as $k\rightarrow \infty$.
Therefore for the zeros of the zeta function we get $\gamma_{k}\sim\,k/\ln k$.

The Weyl theorem and its generalization by  Pleijel relate the asymptotic distribution of eigenvalues of an elliptic differential
operator with geometric parameters of the
surface where the fields satisfy some boundary condition \cite{weyl,pleijel,kac,cou,can,protter}. The asymptotic series for the density of eigenvalues
of the Laplacian operator in a four dimensional space-time is given by $N(\omega)= V\omega^{2}/2\pi^{2}\mp\ S\omega/8\pi+ S{q}/6\pi^2+O(\omega^{-2})$,
where $V$ is the volume of the three-space.
$S$ the area of the boundary and ${q}$ is mean curvature of the boundary averaged over the surface for Dirichlet (Neumann) boundary conditions.
Since the asymptotic behavior of the non-trivial zeros has a regime quite different from the asymptotic behavior of the spectrum of the Laplacian,
we consider a linear operator $-\Delta_{d-1} - {M}$, where
$\Delta_{d-1}$ is the (d-1)-dimensional Laplacian, and impose that the eigenvalues associated with the linear operator ${M}$ are the
imaginary part of the complex zeros of the zeta function.

Suppose a $d+1$-dimensional Minkowski space-time. Let us consider a quantum field theory of a single scalar field $\varphi:\cal M\rightarrow \mathbb{R}$.
The action functional of the theory is
\begin{equation}
S(\varphi)=\int\,d^{d+1}x \,{\cal{L}}(\varphi),
\end{equation}
where ${\cal{L}}(\varphi)$ is the Lagrangian density of the system.
We should assume that the scalar field is coupled to a background field $\sigma(\textbf{x}_\perp,z)$ such
that the system is confined in the interval $[0,a]$ in one dimension and
is unrestricted in the other spatial directions. The Lagrangian density of the system is given by
\begin{eqnarray}
{\cal{L}} =\, &&\varphi(t,\textbf{x}_\perp,z)\bigl(\partial^{2}_t-\Delta_{d-1}-{ M}\bigr)\varphi(t,\textbf{x}_\perp,z)\nonumber\\
&&
-m^{2}\varphi^{2}(t,\textbf{x}_\perp,z)
-\delta m^{2}\varphi^{2}(t,\textbf{x}_\perp,z),
\end{eqnarray}
where $m_B^2=m^2+\delta m^2$, and $m^2_B$ is the squared bare mass and $\delta m^2$ is the mass counterterm.
The counterterms have to be included in odd-dimensional space-time as will be shown below.
The eigenfrequencies of the vacuum modes can be found from the equation
\beq
\bigl(-\Delta_{d-1}-{M}\bigr)\varphi(\textbf{x}_\perp,z)=\omega^2\varphi(\textbf{x}_\perp,z).
\eeq
Since we are assuming that the linear operator ${M}$ has a differential and a background contribution we can write
\beq
-{M}=\bigl(-{\cal O}_{z}+
\sigma(\textbf{x}_\perp,z)\bigr),
\eeq
where ${\cal O}_{z}$ is an unknown differential operator.
It is clear that the linear operator ${-M}$ satisfies
\beq
\bigl(-{\cal O}_{z}+\sigma(\textbf{x}_\perp,z)\bigr)u_{n}(\textbf{x}_\perp,z)=\frac{\gamma_n}{a^2}u_{n}(\textbf{x}_\perp,z),
\eeq
where $u_n(\textbf{x}_\perp,z)$ is a countable infinite set of eigenfunctions. The $u_n(z)$ are normalized eigenfunctions
satisfying the completeness and orthonormality relations, i.e,
\beq
\sum_n u_n(z)u_n^{\ast}(z')= \delta(z-z')
\eeq
and
\beq
\int_0^a\, dz\, u_n(z)u_{n'}^{\ast}(z)= \delta_{n,n'}.
\eeq

The zero-point energy of an massive scalar field defined in a $d$-dimensional box of volume $\Omega$ in a $(d+1)$-dimensional flat space-time is given by
\beq
\langle 0|H|0 \rangle = \frac{1}{2}\sum_{\textbf{k}}^{\infty}\omega_{\textbf{k}}.
\eeq
In the bag configuration, the eigenfrequencies of the vacuum modes are given by
\beq
\omega_{\textbf{k}} = \sqrt{k_1^2 + k_2^2 + \cdots + k_{d-1}^2 + k_d^2+m^{2}},
\eeq
where, as we have discussed,
\beq
k_d(n,a) = \frac{\sqrt{\gamma_n}}{a}\,\,\, n = 1, 2, \cdots,
\eeq
the quantities $\gamma_n$ being the imaginary part of the nontrivial zeros of the Riemann zeta function.
Since the eigenfunctions associated to the $M$ operator are not known, we cannot find the local
renormalized energy density. In other words, local
methods cannot be implemented here.

The zero-point energy
of the scalar field in the slab-bag configuration, taking into account the above equations and using the fact that the integrations over
the momenta correspond to $(d-1)$ integrations and one summation, is thus given by
\begin{eqnarray}
\langle 0|H|0 \rangle =&& \frac{1}{2}\prod_{i = 1}^{d-1}\left(\frac{L_i}{2\pi}\right)\int_{0}^{\infty}d^{d-1}k\times\nonumber\\
&&\sum_{n=1}^{\infty}\left[k_1^2 + \cdots + k_d^2(n,a)+m^{2}\right]^{1/2},
\end{eqnarray}
where $L_i \gg a$. Defining the total zero-point energy per unit area of the hyperplates:
\beq
\varepsilon_{d+1}(a) = \frac{\langle 0|H|0 \rangle}{\prod_{i}L_i},
\eeq
one arrives at
\begin{eqnarray}
\varepsilon_{d+1}(a,m) =&& \frac{1}{2}\sum_{n=1}^{\infty}\int_{0}^{\infty}\,\frac{d^{d-1}k}{(2\pi)^{d-1}}\times\nonumber\\
&&
\left[k_1^2 + \cdots + k_d^2(n,a)+m^{2}\right]^{1/2}.
\label{13}
\end{eqnarray}
In the following we use dimensional \cite{dim1,dim2,dim3} and analytic regularization combined.
Using the formula
\beq
\int_{0}^{\infty}\frac{d^{d}u}{(u^{2}+a^{2})^{s}}=\frac{\pi^{d/2}}{\Gamma(s)}
\Gamma\biggl(s-\frac{d}{2}\biggr)
\frac{1}{(a^{2})^{s-d/2}},
\eeq
and defining the function $f(d)$ as
\beq
f(d)= \frac{1}{2(2\sqrt{\pi})^{d}}
\eeq
the vacuum-energy per unit area $\varepsilon_{d+1}(a)$ can be written as
\beq
\varepsilon_{d+1}(a,m)=\frac{f(d)}{a^{d}}
\Gamma(-d/2)\sum_{n=1}^{\infty}\bigl(\gamma_{n}+a^{2}m^{2}\bigr)^{\frac{d}{2}}.
\label{fun}
\eeq

Due to the unboundedness of the eigenvalues of ${M}$, the vacuum-energy per unit area is divergent in a space-time of any dimension.
In the next section we will discuss the renormalization procedure to obtain a finite vacuum energy associated to the system.

\section{The renormalized vacuum expectation value of the energy operator}

In order to obtain the renormalized vacuum energy, let us use the analytic extension technique.
Before continuing, we would like to point out that a very common procedure to study the divergent contribution of the vacuum energy given by Eq. (\ref{fun}) is to use the
heat-kernel expansion in some asymptotic limit. This is a good procedure for usual operators, such as the Laplacian, where appears a logarithmic divergent mass term
$\ln(4\mu^{2}/m^{2})$. The $\mu$ is an parameter with mass dimension that we have to introduce in order to perform analytic regularizations.
Since in our problem
we are using an unknown operator, this procedure cannot be implemented. In  Appendix B we show that the renormalized vacuum energy is
identified with the constant term in the asymptotic expansion of the regularized vacuum energy. To proceed,
let us assume that $a^{2}m^{2}<\gamma_{1}$. In other words, the square of the ratio between the length of the slab-bag in the
$z$ direction to the Compton length $l_{c}=m^{-1}$ is smaller than
the  first non-trivial zero.
It is important to stress that in general it is not necessary to impose such condition. Here we need this assumption to make
use of the generalization of the binomial expansion.
For any complex coefficient $\alpha$ and $|x|<1$ we may write
\begin{equation}
(1+x)^{\alpha}= \sum_{n=0}^{N}C_{\alpha}^{n}\, x^n,
\end{equation}
where $C_{\alpha}^{n}$ are the generalizations of the binomial coefficients given by $C_{\alpha}^{n}=
\frac{\Gamma(\alpha+1)}{\Gamma(n+1)\Gamma(\alpha-n +1)}$. For odd dimensional
space-times we get that the sum has a finite number of terms, i.e., $N=d/2$ and for even dimensional space-time we have $N\rightarrow\infty$.
Let us define the function $g(k,d)$ as
\begin{equation}
g(d,k)=\frac{f(d)\Gamma(\frac{d}{2}+1)}{\Gamma(k+1)\Gamma\bigr(\frac{d}{2}-k+1\bigl)}.
\label{g}
\end{equation}
The zero-point energy per unit area can be written as
\begin{eqnarray}
\varepsilon_{d+1}(a,m)=&&\frac{1}{a^{d}}\sum_{k=0}^{N(d)}g(d,k)\Gamma(-d/2)(m a)^{2k}
\times\nonumber\\
&&\sum_{n =1}^{\infty}(\gamma_{n})^{\frac{d}{2}-k}.
\end{eqnarray}
Let us define the super-zeta or secondary zeta function built over the Riemann zeros, i.e.,
the nontrivial zeros of the Riemann zeta function. Let $s$ be a complex variable i.e. $s=\sigma+i\tau$ with $\sigma,\tau \in \mathbb{R}$.
The super-zeta $G_{\gamma}(s)$ is defined as
\begin{equation}
G_{\gamma}(s)=\sum_{n =1}^{\infty}\frac{1}{\gamma_{n}^{s}}, \,\,\,\Re(s)>1,
\label{zetazero}
\end{equation}
where we are assuming that $\gamma_{n}>0$.
The analytic continuation of
the super-zetas has been discussed in the literature ~\cite{gui,delsarte,cha1,cha2,ivic,voros1}.
In the Appendix A, we present the analytic extension for $G_{\gamma}(s)$ assuming the Riemann hypothesis and following the Ref.~\cite{delsarte}.
A more detailed study for the zeta function for the Riemann zeros
can be found in Ref.~\cite{superzeta}.

Let us study first the even dimensional space-time case. By using the definition of $g(d,k)$ of Eq. (\ref{g}) we define the function
%
$h(d,k)=g(d,k)\Gamma(-d/2)$.
%
The zero-point energy per unit area can be written as
\begin{eqnarray}
\varepsilon_{d+1}(a,m)=\frac{1}{a^d}\sum_{k=0}^{\infty}h(d,k)(ma)^{2k}G_{\gamma}\bigr(k-d/2\bigl).
\label{even-energy}
\end{eqnarray}
By the use of the analytic extension of the super-zeta $G_{\gamma}(s)$ we obtain that the regularized
vacuum energy of the system is finite in any even dimensional space-time.

The zero-point energy per unit area $\varepsilon_{d+1}(a)$ in the odd dimensional space-time case can be written as
\begin{eqnarray}
\varepsilon_{d+1}(a,m)=&&\frac{1}{a^d}\sum_{k=0}^{d/2}g(d,k)(ma)^{2k}\times\nonumber\\
&&
\Gamma(-d/2)\,G_{\gamma}\bigr(k-d/2\bigl).
\end{eqnarray}
The analytic continuation of $G_{\gamma}(s)$ has simple poles for $(4n+3)$ dimensional space-time, for $n=0,1,2,..$. Since the gamma function has poles
for negative integers, we conclude that to find a finite vacuum energy in odd-dimensional space-time, we have to introduce
mass counterterms.
The gamma function $\Gamma(z)$ is
a meromorphic function of a complex variable $z$ with simple poles
at the points $z=0,-1,-2..\,$. In the neighborhood of any of its poles $z=-n$ for $n=0,1,2..$ the gamma function has a representation given by
\begin{equation}
\Gamma(z)=\frac{(-1)^{n}}{n!(z+n)}+\Omega(z+n)
\end{equation}
where $\Omega(z+n)$ stands for the regular part of the analytic extension. The same construction can be used for the super-zeta function.
In the neighborhood of any of its poles $z=-(2m+1)$ for $m=0,1,2...$ the super-zeta function has a representation given by
\begin{equation}
G_{\gamma}(z)=\frac{\alpha_{1}}{z+(2m+1)}+\Phi\bigl(z+(2m+1)\bigr),
\end{equation}
where $\Phi(z+(2m+1))$ is the regular part of the analytic extension and $\alpha_1$ is a constant.
The regularized vacuum energy has
second order and first order poles for $(4n+3)$-dimensional space-time with $n=0,1,2..\,$. Just first order poles for odd-dimensional space-time
excluding the $(4n+3)$-dimensional poles, and a computable finite part which would be the renormalized vacuum energy. The term $k=0$ does not depend on the mass.
Therefore in order to make the energy density per unit area finite in odd dimensional space-time we introduce first mass counterterms (associated to this $k=0$ term)
proportional to second order and first order poles. The other terms $k> 0$ which depend on the mass give us first order poles.
To render the density energy per unit area finite we introduce also mass counterterms proportional to these first order poles.
This problem has been discussed by Kay \cite{kay}. This author
shows that the analytic regularization procedure does not yield a finite result automatically in the case of massive fields.

Let us discuss this renormalization procedure in a three-dimensional space-time. In this case we have
\begin{eqnarray}
\varepsilon_{3}(a,m)&=&\frac{1}{a^2}\sum_{k=0}^{1}g(2,k)(ma)^{2k}\Gamma(-1+\epsilon)\times\nonumber\\
&& G_{\gamma}\bigr(k-1+\epsilon\,\bigl).
\end{eqnarray}
The mass counterterm $\delta m^2$ in a generic odd-dimensional space-time has dimension $[\delta m^{2}]=\frac{1}{a^{d+1}}\mu^{1-d}$,
where we introduced a mass parameter $\mu$. Therefore we get
\begin{eqnarray}
\delta m^2=&&-\frac{1}{8\pi\,a^3\mu}\biggr(\frac{1}{\epsilon^2}+\frac{1}{\epsilon}(\Omega_1 +\Phi_1)\biggl)+\nonumber\\
&&-\frac{1}{8\pi\,a\mu}\frac{1}{\epsilon}m^2 G_{\gamma}(0),
\end{eqnarray}
where $\Omega_{1}$ is the regular part of the analytic extension for the gamma function around the pole in $z=-1$ and $\Phi_1$ is
the regular part of the analytic extension for the super-zeta
function around the pole in $z=-1$.

The mass correction $\Delta m^2$ is given by
\begin{eqnarray}
\Delta m^2(a,\mu)&=&\frac{1}{8\pi}\biggl(\frac{1}{a^3\mu}\Omega_1\Phi_1+
\frac{1}{a\mu}m^2\Omega_1G_{\gamma}(0)\biggr),
\end{eqnarray}
where $\mu$ is a mass parameter that has take to introduced to perform the analytic regularization.
We can define a renormalized square mass, the Riemann mass, given by
\begin{eqnarray}
m_{R}^{2}=m^{2}+\Delta m^2(a,\mu).
\end{eqnarray}

It is interesting to stress that the mass generation occurs only
in odd dimensional space-time, and that the effect vanishes when the translational invariance is recovered. This mechanism is quite similar with the topological
generation of mass \cite{yo, ford}. In this case, the translational invariance is maintained.
\section{Conclusions}

"Do you know  a physical reason that the Riemann hypothesis should be true?"(E. Landau). Our knowledge of number theory and related areas is not powerful enough to
prove it, until now. The aim of this paper is to investigate the consequences
in the quantum field theory framework of accepting the truth of this mathematical proposition.

From Hadamard's theory it is possible to define analytic functions by its zeros and singularities. In the case of the Riemann zeta function, it is possible
to represent it as an Euler product and a Hadamard's product. This shows a reciprocity between the set of prime numbers and nontrivial zeros of the zeta function.
Nevertheless these two sequences of numbers have totally distinct behavior with respect to being the spectrum of a
linear operator associated to a system with countable infinite number of degrees of freedom.

In two recent papers it was investigated whether number-theoretical sequences can be associated with the spectrum of some
hypothetical linear operator of physical systems with infinite number of degrees of freedom.
In the first one it
was shown that the sequence of prime numbers is not zeta
regularizable~\cite{ga}, therefore this sequence of numbers
cannot be the spectrum of a linear operator described by functional integrals. In the second one \cite{gn2} it was shown that it is possible to extend the
Hilbert-P\'olya conjecture to systems with with countably infinite number of degrees of freedom described by functional integrals.

In this paper, the renormalized zero-point energy of a massive scalar field with the Riemann zeros in the spectrum of the vacuum modes is presented. Using analytic and dimensional regularization,
for even dimensional space-time, we show that the series that defines the regularized energy density is
finite. For odd dimensional space-time the analytic regularization procedure does not produce finite results because the coefficients of the series are divergent.
We concluded that in order to renormalize the vacuum energy one is forced to introduce mass counterterms in the
interaction Lagrangian for any odd dimensional space-time.

Finally two comments are in order. First, the renormalized vacuum energy is identified with the constant term in the asymptotic expansion of the regularized vacuum energy.
For the massless case it is possible to show, using an exponential ultraviolet cut-off~\cite{kay,ruggiero,ss,nami1,nami2,bene}, that the finite part of the regularized vacuum energy
agrees with the renormalized energy density obtained by the analytic regularization procedure.
Second, the Casimir energy for the case of fermionic fields was investigated in this configuration,
based on the MIT bag model \cite{mit1,mit2}. The generalization for massive fermions was presented by Elizalde et al \cite{el}.
See also the Ref. \cite{oi}. The study of fermionic fields where the vacuum modes contain in the spectrum the nontrivial zeros of the Riemann zeta function
is under investigation by the authors.

\section{Acknowlegements}

We would like to thanks Martin Makler, Benar Svaiter, Jorge Stephany Ruiz and Gabriel Menezes for useful discussions.
This paper was supported by Conselho Nacional de Desenvolvimento
Cientifico e Tecnol{\'o}gico do Brazil (CNPq).

\begin{appendix}
\makeatletter \@addtoreset{equation}{section} \makeatother
\renewcommand{\theequation}{\thesection.\arabic{equation}}

\section{The analytic extension of the super-zeta function $G_{\gamma}(s)$}

In this appendix we present the analytic extension of one of the super-zeta functions.
We are following the Ref.~\cite{delsarte}. Using the definition given by Eq.~(\ref{zetazero}) we get
\begin{equation}
\Gamma\biggl(\frac{s}{2}\biggr)\pi^{-\frac{s}{2}}G_{\gamma}(s)=\int_{0}^{\infty}\,dx\,
x^{\frac{s}{2}-1}\sum_{\gamma>0}\,e^{-\pi\gamma^{2}x}.
\label{zetazero2}
\end{equation}
Let us split the integral that appears in Eq.~(\ref{zetazero2}) in the intervals $[0,1]$ and $[1,\infty)$, and define the functions
\begin{equation}
A(s)=\int_{0}^{1}\,dx\, x^{\frac{s}{2}-1}\sum_{\gamma>0}\,e^{-\pi\gamma^{2}x}
\label{zetazero3}
\end{equation}
and
\begin{equation}
B(s)=\int_{1}^{\infty}\,dx\, x^{\frac{s}{2}-1}\sum_{\gamma>0}\,e^{-\pi\gamma^{2}x}.
\label{zetazero4}
\end{equation}
Note that $B(s)$ is an entire function. To proceed let us use that
\begin{eqnarray}
\sum_{\gamma>0}\,e^{-\pi\gamma^{2}x} = &&-\frac{1}{2\pi\sqrt{x}}\sum_{n=2}^{\infty}
\frac{\Lambda(n)}{\sqrt{n}}\,e^{-\frac{(\ln n)^{2}}{4\pi x}}+\,e^{\frac{\pi x}{4}}\nonumber\\
&&-\frac{1}{2\pi}\int_{0}^{\infty}dt\,e^{-\pi x t^{2}}\Psi(t),
\label{zetazero55}
\end{eqnarray}
where the function $\Psi(t)$ is given by
\begin{equation}
\Psi(t)=\frac{\zeta'(\frac{1}{2}+i t)}{\zeta(\frac{1}{2}+i t)}+\frac{\zeta'(\frac{1}{2}-i t)}{\zeta(\frac{1}{2}-i t)}.
\label{zetazero5}
\end{equation}
Substituting Eq.~(\ref{zetazero55}) in~(\ref{zetazero3}) we get that $A$-function can be written as
\begin{equation}
A(s)=A_{1}(s)+A_{2}(s)+A_{3}(s),
\label{zetazero6}
\end{equation}
where
\begin{equation}
A_{1}(s)=-\frac{1}{2\pi}\int_{0}^{1}\,dx\,x^{\frac{s}{2}-\frac{3}{2}}\biggl(\sum_{n=2}^{\infty}
\frac{\Lambda(n)}{\sqrt{n}}\,e^{-\frac{(\ln n)^{2}}{4\pi x}}\biggr),
\label{zetazero7}
\end{equation}
\begin{equation}
A_{2}(s)=\int_{0}^{1}\,dx\,x^{\frac{s}{2}-1}\,e^{\frac{\pi x}{4}}
\label{zetazero8}
\end{equation}
and finally
\begin{equation}
A_{3}(s)=-\frac{1}{2\pi}\int_{0}^{1}\,dx\,x^{\frac{s}{2}-1}\,\biggl(\int_{0}^{\infty}\,e^{-\pi x t^{2}}\Psi(t)\biggr).
\label{zetazero9}
\end{equation}
Changing variables in the $A_{1}(s)$, i.e., $x\rightarrow1/x$ we get
\begin{equation}
A_{1}(s)=-\frac{1}{2\pi}\int_{1}^{\infty}\,dx\,x^{-\frac{s}{2}-\frac{1}{2}}
\biggl(\sum_{n=2}^{\infty}\frac{\Lambda(n)}{\sqrt{n}}\,e^{-\frac{x(\ln n)^{2}}{4\pi}}\biggr).
\label{zetazero10}
\end{equation}
It is clear that $A_{1}(s)$ is an entire function of $s$. Let us define $\Phi(s)$ as
\begin{equation}
\Phi(s)=A_{1}(s)+B(s).
\label{zetazero11}
\end{equation}
Using Eqs.~(\ref{zetazero4}),~(\ref{zetazero6}),~(\ref{zetazero8}),~(\ref{zetazero9}) and~(\ref{zetazero11}) we can write expression~(\ref{zetazero2}) as
\begin{equation}
\Gamma\biggl(\frac{s}{2}\biggr)\pi^{-\frac{s}{2}}G_{\gamma}(s)=\Phi(s)+A_{2}(s)+A_{3}(s).
\label{zetazero12}
\end{equation}
Since $\Phi(s)$ is an entire function and we have the integrals that define $A_{2}(s)$ and $A_{3}(s)$, the above formula is
the analytic extension of the secondary zeta function.
The function
$G_{\gamma}(s)$ is a meromorphic function of $s$ in the whole complex plane with double pole at
$s=1$ and simple poles at $s=-1,-3,..,-(2n+1),..\,$. Therefore $(s-1)^{2}G_{\gamma}(s)(\Gamma(s))^{-1}$ is an entire function.

\section{Asymptotic expansion of the regularized vacuum energy}

Since the renormalized vacuum energy is identified with the constant term in the asymptotic expansion of the regularized vacuum energy
the aim of this appendix is to discuss the divergent and the finite contribution for the vacuum energy using the cut-off method. For simplicity we use $m=0$.
To proceed, the angular part of the $(d-1)$-dimensional integral of Eq.~(\ref{13}) can be easily calculated with the aid of the result
\begin{equation}
\int d^{d-1}k=\frac{2\pi^\frac{d-1}{2}}{\Gamma(\frac{d-1}{2})}\int_{0}^{\infty}\,dr\,r^{d-2},
\end{equation}
where $r = \sqrt{k_1^2 + \cdots + k_{d-1}^2}$. Defining the function $F(d)$ as
\begin{equation}
F(d)= \frac{1}{(2\sqrt{\pi})^{d-1}}\,\frac{1}{\Gamma(\frac{d-1}{2})},
\end{equation}
one can rewrite the vacuum energy per unit area as
\beq
\varepsilon_{d+1}(a) = F(d)\sum_{n=1}^{\infty}\int_{0}^{\infty}\,dr\,r^{d-2}\left[r^2 + k_d^2(n,a)\right]^{1/2}
\eeq
Using that $k_d(n,a) = \sqrt{\gamma_n}/a$, and simple algebraic manipulations allow us to write the vacuum energy per unit area $\varepsilon_{d+1}(a)$ as
\begin{equation}
\varepsilon_{d+1}(a) =  \frac{F(d)}{2a^{d}}\int_{0}^{\infty}dv\,v^{(d-3)/2}\sum_{n=1}^{\infty}\left(v+\gamma_{n}\right)^{1/2}.
\end{equation}
As usual, let us introduce an ultraviolet cut-off given by
\beq
e^{-\lambda\left(v+\gamma_{n}\right)^{1/2}},\,\,\,\Re(\lambda) > 0.
\eeq
The regularized vacuum energy per unit area $\varepsilon_{d+1}(a,\lambda)$ is then given by
\begin{eqnarray}
\varepsilon_{d+1}(a,\lambda) = &&\frac{F(d)}{2a^{d}}\int_{0}^{\infty}dv\,v^{(d-3)/2}\sum_{n=1}^{\infty}
\left(v+\gamma_{n}\right)^{1/2}\,\nonumber\\
&& \times e^{-\lambda\left(v+\gamma_{n}\right)^{1/2}}.
\end{eqnarray}
After some algebra one gets
\begin{eqnarray}
\varepsilon_{d+1}(a,\lambda) = && \frac{F(d)}{a^{d}}
\sum_{n=1}^{\infty}
\int_{\sqrt{\gamma_n}}^{\infty} du\, u^{d-1}\left(1-\frac{\gamma_n}{u^2}\right)^{(d-3)/2}\nonumber\\
&& \times e^{-\lambda u}.
\end{eqnarray}
Here we use again the generalization for the binomial expansion for any real coefficient $\alpha$ and $|x|<1$.
For even dimensional
space-time we get that the sum has a finite number of terms, i.e. $N=\frac{d-3}{2}$ and for odd dimensional space-time we have infinite terms.
From section III, we know that only for even dimensional space-time, the renormalized vacuum energy is finite, without
the necessity of introducing counterterms.
Let us discuss the even dimensional case. After some algebra we can write the asymptotic expansion in the form
\begin{eqnarray}
\varepsilon_{d+1}(a,\lambda) = && \frac{1}{a^d}\sum_{k=0}^{\frac{d-3}{2}}g_1(d,k)
\sum_{n=1}^{\infty}(-1)^{k}\gamma_{n}^{k}\times\nonumber\\
&&\int_{\sqrt{\gamma_n}}^{\infty} du\, u^{(d-2k)-1}\,e^{-\lambda u},
\end{eqnarray}
where $g_1(d,k)=\frac{4\sqrt\pi\,F(d)}{\Gamma(k+1)\Gamma\left(\frac{d-1-2k}{2}\right)}$.
In order to proceed let use the definition of the incomplete gamma function $\Gamma(a,z)$\cite{abram}
\beq
\Gamma(a,z)=\int_{z}^{\infty}t^{a-1}\,e^{-t}\,dt,
\eeq
which has the following series expansion \cite{grads,ederlyi}
\begin{eqnarray}
\Gamma(a,z)\,&&=\,\Gamma(a)\left[1-z^a\,e^{-z}\sum_{l=0}^{\infty}\frac{z^l}{\Gamma(a+l+1)}\right], \nonumber\\
&& a\neq 0,-1,-2...
\end{eqnarray}
Defining $h_1(d,k)=g_1(d,k)\Gamma(d-2k)$, the density energy therefore may be written as
\begin{eqnarray}
&&\varepsilon_{d+1}(a,\lambda)=\frac{1}{a^d}\sum_{k=0}^{\frac{d-3}{2}}h_1(d,k)\sum_{n=1}^{\infty}
\frac{(-1)^{k}\gamma_n^{k}}{\lambda^{d-2k}}\times\nonumber\\
&&\biggl[1-(\lambda\sqrt{\gamma_n})^{d-2k}
\sum_{l=0}^{\infty}\frac{\,e^{-\lambda\sqrt{\gamma_n}}(\lambda\sqrt{\gamma_n})^l}{\Gamma(d-2k+l+1)}\biggr].
\end{eqnarray}
Let us study the polar part of the regularized energy density. We get
\begin{eqnarray}
&&\varepsilon_{d+1}(a,\lambda)=\frac{1}{a^d}\sum_{k=0}^{\frac{d-3}{2}}(-1)^{k}h_1(d,k) \nonumber\\
&&\times \sum_{n=1}^{\infty}\biggl[
\frac{\gamma_{n}^{-k}}{\lambda^{d-2k}} - \frac{\gamma_n^{-\frac{d}{2}}-\lambda\gamma_{n}^{-\frac{d+1}{2}}}{\Gamma(d-2k+1)}
+...\biggr]
\end{eqnarray}
where a term of order $\lambda^2$ was neglected. Note that the cut-off regularization is not enough to obtain a finite result for the vacuum energy.
We also need to apply an analytic regularization procedure.
To see this let us investigate the particular case of $d=3$, which implies $k=0$. We have that $\varepsilon_{4}(a,\lambda)$ can be written as
\beq
\varepsilon_{4}(a,\lambda)=-\frac{1}{\pi a^3}\left[\frac{2}{\lambda^3}G_{\gamma}(0) -\frac{1}{3}G_{\gamma}(-3/2) +
\frac{5\lambda}{12}\,G_{\gamma}(-2)\right].
\eeq
The first term in the right hand side in the above equation is divergent when the cut-off is removed. The second term is finite by the use of analytic regularization procedure.
This term coincides with the density energy for the system using an analytic regularization procedure. Since
$G_{\gamma}(-2)$ is regular, the third one goes to zero when the cut-off is removed.
 The finite part gives us the correct renormalized energy density. It expresses the fact that the
 renormalized vacuum energy is identified with the constant term in the asymptotic expansion of the regularized vacuum energy \cite{ful}.

\end{appendix}

\end{document}